\documentclass[%
 reprint,
 amsmath,amssymb,
 aps,
 superscriptaddress,
]{revtex4-2}

\usepackage{makecell}
\usepackage{graphicx}
\usepackage{dcolumn}
\usepackage{bm}
\usepackage{xcolor} 
\usepackage{upgreek}

\begin{document}

\preprint{APS/123-QED}
\title{Enhanced sliding in the coexisting Charge Density Wave phase of strained TbTe$_3$}

\author{A. Gallo--Frantz}
\affiliation{Laboratoire de Physique des Solides, Université Paris-Saclay, CNRS, 91405, Orsay, France}
\author{A.A. Sinchenko}
\affiliation{Laboratoire de Physique des Solides, Université Paris-Saclay, CNRS, 91405, Orsay, France}
\affiliation{Kotelnikov Institute of Radioengineering and Electronics of RAS, 125009 Moscow, Russia}
\author{L. Ortega}
\affiliation{Laboratoire de Physique des Solides, Université Paris-Saclay, CNRS, 91405, Orsay, France}
\author{A. Hadj-Azzem}
\author{J.E Lorenzo}
\author{P. Monceau}
\affiliation{Univ. Grenoble Alpes, CNRS, Grenoble INP, Institut Néel, 38000, Grenoble, France}
\author{D. Le Bolloc'h}
\affiliation{Laboratoire de Physique des Solides, Université Paris-Saclay, CNRS, 91405, Orsay, France}
\author{V.L.R Jacques}
\affiliation{Laboratoire de Physique des Solides, Université Paris-Saclay, CNRS, 91405, Orsay, France}

\date{\today}

\begin{abstract}
In the last years, the application of mechanical tensile stress has been shown to trigger unconventional phases in quantum materials. Recently, an orientational transition of the charge density wave (CDW) of TbTe$_3$ was reported under the application of moderate uniaxial tensile stress, associated to the in-plane a/c lattice anisotropy. This leads to the disappearance of the pristine c-oriented CDW (CDW$_c$) and to the appearance of the orthogonal a-oriented one (CDW$_a$). A coexistence phase with a superposition of CDW$_a$ and CDW$_c$ appears when a=c. Here, we probe the sliding dynamics of both CDW$_c$ and CDW$_a$ as a function of applied tensile stress, \textit{i.e.} in the pure CDW$_c$, pure CDW$_a$ and coexistence phases, by measuring current-voltage characteristics and associated differential resistances. If the pristine CDW$_c$ has been known to display non-linear dynamics and thus sliding for several years, we show here that the strain-induced CDW$_a$ phase also displays the same characteristics, proving its incommensurate nature. Moreover, we show that sliding also takes place both for CDW$_c$ and CDW$_a$, with reduced threshold currents, in the coexistence phase, which questions the microstructure of these coexisting orthogonal CDWs in real space. Finally, the linearity of the threshold fields is kept for all deformation states, with finite threshold fields at T$_c$, which is discussed in terms of commensurability and pinning potentials. This work opens new questions for the theoretical description of sliding CDWs in quasi-2D systems.
\end{abstract}

\maketitle

\section{\label{sec:Introduction} Introduction}

The charge density wave (CDW) is an electronic instability appearing in low-dimensional systems below a critical temperature \cite{monceau_electronic_2012}. It is characterized by a simultaneous periodic lattice distortion and a modulation of the electronic density $\rho(x) \sim cos(2.k_F.x + \phi)$ that induces the opening of a gap $\Delta$ in the electronic spectrum. This electronic quantum state can be described by the order parameter $\Psi \sim  \Delta e^{i\phi}$. Its complex nature allows for collective excitations: amplitude ($\Delta$) and phase ($\phi$) modes \cite{gruner_density_2019}. One of the most interesting properties of incommensurate CDW systems is its collective motion, also called sliding. Predicted as dissipationless electron transport by Fröhlich \cite{frohlich_theory_1954}, sliding has been explained as a time-dependent and space-independent phase mode $\phi(t)$ in an ideal unidimensional system. However, in real systems, several mechanisms pin the phase $\phi$ such as bulk or surface defects \cite{gruner_density_2019, bellec_evidence_2020, bellec_essential_2020}, impurities \cite{maki_thermal_1986, maki_impurity_1989, maki_phase_1990}, interactions \cite{sinchenko_dynamical_2013} or commensurability with the main lattice. Then, the ground state is the result of the competition between the pinning mechanisms and the elastic deformations of the CDW \cite{le_bolloch_effect_2016, bellec_evidence_2020, bellec_essential_2020, le_bolloch_importance_2025}. To initiate CDW sliding, the application of an external electric field larger than a characteristic threshold $E_{t}$, determined by pinning, is required \cite{gruner_density_2019}. Sliding is manifested as a sharp increase of conductivity and then a sharp drop in the differential resistance. Such transport properties have been observed in various quasi-on-dimensional (quasi-1D) inorganic and organic compounds \cite{monceau_electronic_2012}.\\

\begin{figure}[h]
    \centering
    \includegraphics[width = 1.00\linewidth]{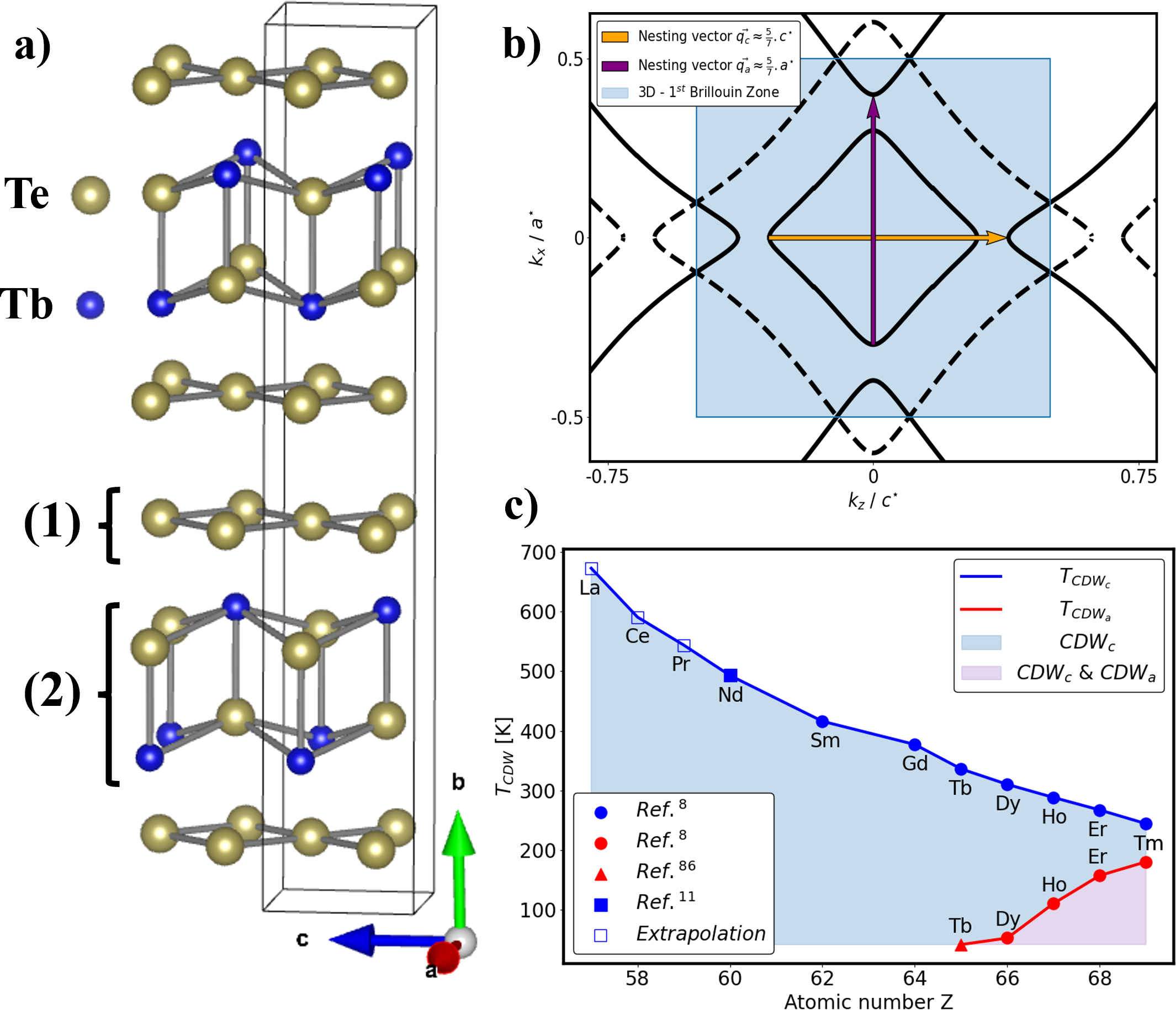}
    \caption{a) $TbTe_3$ crystal structure made of almost square $Te$ layers (1) and $RTe$ slabs (2). b) Sketch of the Fermi surface. Nesting conditions (orange and purple arrows) inducing CDWs formations (See Ref.\cite{brouet_fermi_2004, brouet_angle-resolved_2008, moore_fermi_2010} for Fermi surface reconstruction). c) Temperature - Rare-earth element phase-digram of $RTe_3$. All $RTe_3$ exhibit $CDW_c$ (blue line and dots) while only heavier R elements ($Tb, Dy, Ho, Er, Tm$) exhibit $CDW_a$ coexisting with the $CDW_c$ (red line and dots) due to nesting conditions \cite{malliakas_divergence_2006, ru_effect_2008, banerjee_charge_2013}.}
    \label{fig:TbTe3_crystal_structure}
\end{figure}

\noindent Recently, CDW sliding has also been observed in the family of quasi-two-dimensional (quasi-2D) rare-earth tritellurides $RTe_{3}$ ($R = La, Ce, Pr, Nd, Gd, Tb, Dy, Er, Tm$) \cite{sinchenko_sliding_2012, sinchenko_unidirectional_2014, sinchenko_dynamical_2016}. In the last two decades, these compounds have raised an intense research activity due to the competing electronic orders in their rich phase diagrams probed by several experimental techniques, as well as their sensitivity under various stimuli such as chemical substitution \cite{dimasi_chemical_1995, brouet_fermi_2004, malliakas_divergence_2006, ru_effect_2008, ru_magnetic_2008, brouet_angle-resolved_2008, tomic_scanning_2009, moore_fermi_2010, banerjee_charge_2013, sinchenko_spontaneous_2014, hu_coexistence_2014, raghavan_atomic-scale_2024, yumigeta_alloying_2024}, Palladium doping \cite{lou_interplay_2016, straquadine_suppression_2019, mallayya_bragg_2024, singh_effect_2025}, defects \cite{fang_stm_2007, fu_multiple_2016, siddique_realignment_2024}, hydrostatic pressure \cite{sacchetti_pressure_2007, hamlin_pressure-induced_2009, sacchetti_pressure-induced_2009, zocco_high-pressure_2009, zocco_pressure_2015, kopaczek_pressure-induced_2022}, uniaxial and biaxial tensile stresses \cite{straquadine_evidence_2022, gallofrantz_charge_2024, singh_emergent_2024, freitas_revealing_2026}, light pulses \cite{chen_revealing_2014, zong_evidence_2019, kogar_light-induced_2020, gonzalez-vallejo_time-resolved_2022, kim_emergent_2024, dellangela_time_2025}, magnetic field \cite{iyeiri_magnetic_2003, deguchi_magnetic_2009, frolov_magnetoresistance_2018, volkova_magnetic_2022, zaitsev-zotov_slow_2025}, electric field \cite{sinchenko_sliding_2012, sinchenko_unidirectional_2014, sinchenko_dynamical_2016, le_bolloch_effect_2016} or aging \cite{frolov_features_2019, frolov_toward_2020, frolov_non-equilibrium_2021, frolov_logarithmic_2023}. See Ref.\cite{yumigeta_advances_2021} for a review. These layered compounds show a weakly orthorhombic crystal structure (space group : $Cmcm$) and are made of double $Te$ square layers spaced by $RTe$ slabs. They are stacked and linked by Van-der-Waals bonds along the long b-axis which is perpendicular to the $Te$ planes. The a- and c-axis, lying in the $Te$ plane, are practically equal ($a/c \sim 0.998 < 1$) making the compounds quasi-tetragonal (nearly square $Te$-plane) and inducing almost isotropic electronic properties in the (a-c) plane \cite{brouet_angle-resolved_2008, moore_fermi_2010, sinchenko_spontaneous_2014}. Their quasi-2D crystalline structure, dominated by p-orbital character gives rise to slightly distorted electronic bands and strong Fermi-surface nesting conditions. These nesting features, which are relatively well captured within a tight-binding description, can provide the conditions for the emergence of CDW phases. The crystal structure and a sketch of the Fermi surface are is illustrated in Fig.\ref{fig:TbTe3_crystal_structure}a) and Fig.\ref{fig:TbTe3_crystal_structure}b), respectively. All compounds of this family exhibit a high-temperature incommensurate CDW with a wavevector $\textbf{q}_c \sim (0, \:\: 0, \:\: 5/7c^{\star})$ ($CDW_c$) while only the heavier R ($Tb, Dy, Ho, Er, Tm$) show an additional second CDW occurring at low-temperature with a wavevector $\textbf{q}_a \sim (2/3 a^{\star}, \:\: 0, \:\: 0)$ ($CDW_a$), perpendicular to $\textbf{q}_c$ as shown in Fig.\ref{fig:TbTe3_crystal_structure}c) \cite{malliakas_divergence_2006, ru_effect_2008, banerjee_charge_2013, maschek_competing_2018}.

\newpage
\noindent The application of tensile stress along a-axis results in an inversion of the $a/c$-ratio (from $a/c < 1$ to $a/c > 1$) together with an orientational transition from c-axis to a-axis for the CDW (from $CDW_c$ to $CDW_a$). Transport and XRD measurements show a region with the coexistence of both CDW orders close to $a/c \sim 1$, and suggest a similar wavevector magnitude different than the low-temperature CDW order ($CDW_c : \textbf{q}_c \sim (0, \:\: 0, \:\: 5/7c^{\star}) \:\: \rightarrow \:\: CDW_a : \textbf{q}_a \sim (5/7a^{\star}, \:\: 0, \:\: 0)$) \cite{gallofrantz_charge_2024}. Similar behaviour is reported in other rare-earth tritellurides ($ErTe_3$, $TmTe_3$) \cite{straquadine_evidence_2022, singh_emergent_2024, kim_emergent_2024}.

\noindent Following of our previous works \cite{sinchenko_unidirectional_2014, gallofrantz_charge_2024}, we report here the strain-temperature dependance of the sliding properties for the high-temperature pristine state $CDW_c$ and the strain-induced $CDW_a$ in $TbTe_3$. These measurements allow to highlight the incommensurability of strain-induced $CDW_a$ with the host lattice but the variation of the electrical threshold as a function of the strain questions how the orders coexist in the crystal.

\newpage
\section{\label{sec:Experimental_methods} Experimental methods}
\subsection{Samples}

\begin{figure}[h]
    \centering
    \includegraphics[width = 1.00\linewidth]{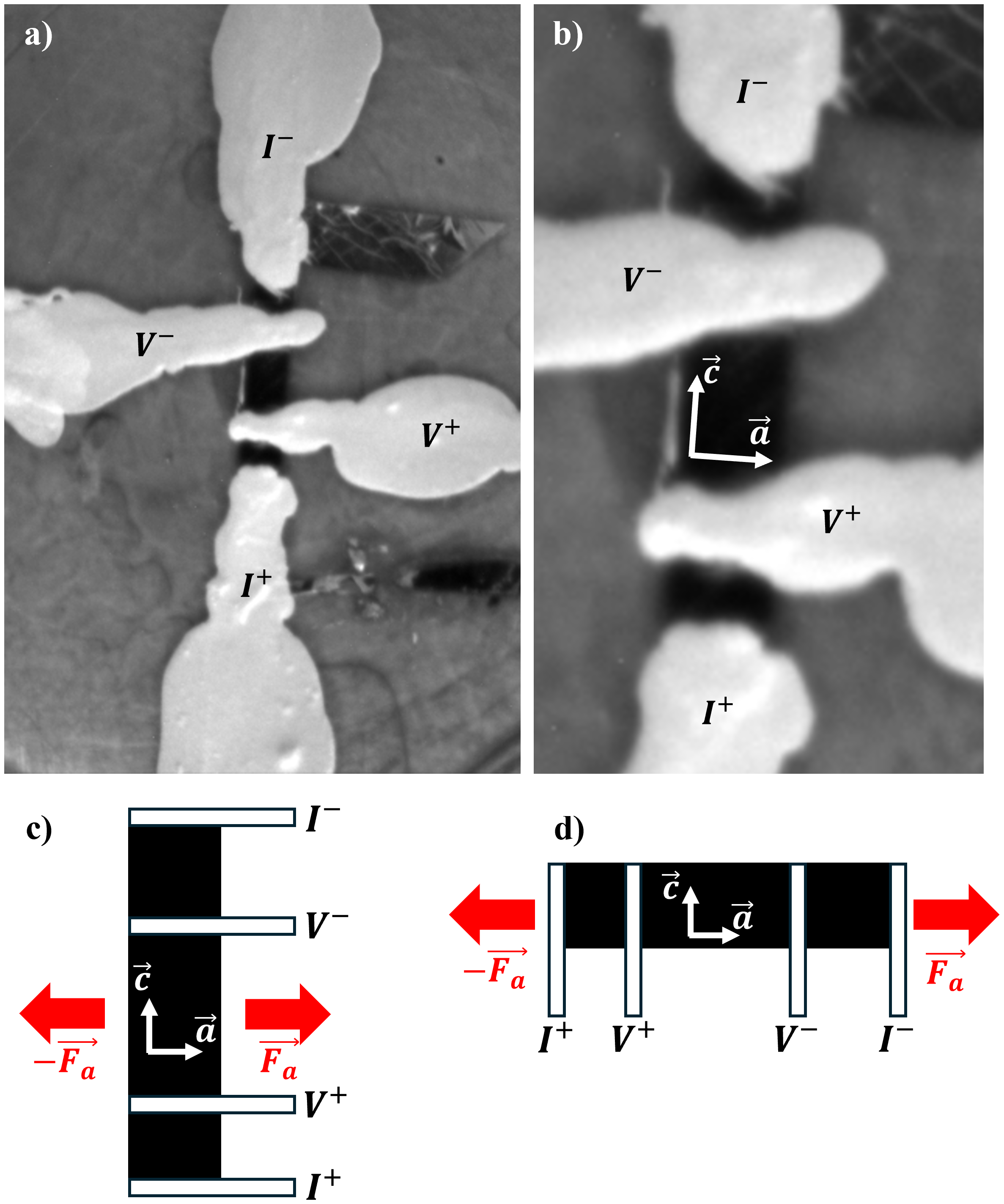}
    \caption{a) Rectangular shaped $TbTe_3$ sample 1 (black rectangle) with electrical contacts (silver paint) in four-bar geometry. b) Zoom on $TbTe_3$ sample 1. c- and a-axis are represented with white arrows. c) and d) are sketches of sample 1 and sample 2, respectively. Crystallographic axis are represented with white arrows while applied tensile stress along a-axis (see Sec.\ref{sec:Experimental_methods}.B) is shown with red arrow for both panels.}
    \label{fig:TbTe3_sample1_picture}
\end{figure}

Single crystals of $TbTe_3$ were grown by the self-flux method, as described in Ref.\cite{sinchenko_sliding_2012}. For these experiments, two samples were selected, cut in rectangular shape and then mechanically exfoliated. They are thinned down to few $\mu m$ to get homogeneous deformation in the samples volume. Sample 1 was cut along c-axis while sample 2 was cut along a-axis to probe the non-linear transport properties of $CDW_c$ and strain-induced $CDW_a$, respectively. Sample 1 is shown in Fig.\ref{fig:TbTe3_sample1_picture}, with its crystallographic orientation and electrical contacts.

\noindent Sample 1 was probed both by X-ray diffraction (XRD) with uniaxial tensile stresses along c-axis to get structural properties and four-probe resistivity measurements with uniaxial tensile stresses along c- and a-axis. Sample 2 was probed only by four-probe resistivity measurement with uniaxial tensile stress along the a-axis.

\newpage
\subsection{Uniaxial tensile stresses}

The device used to apply tensile stress is described in the supplementary information of Ref.\cite{gallofrantz_charge_2024}. The $TbTe_3$ samples were glued on a $125 \: \mu m$-thick polyimide cross-shaped deformable substrate with the crystallographic a- and c-axis aligned with the arms of the cross. The substrate was installed in the biaxial tensile stress device, attached to four independent motors that can pull on each branch independently and mechanically stretched in both in-plane directions. The forces applied along the arms were measured, in Newtons (N), with calibrated force sensors. The center of the cross-shaped substrate lied on the cold finger of a helium-flow Konti-Micro cryostat from CryoVac GmbH, allowing to reach temperatures in the range $15$-$200 \: K$. The whole setup is placed under vacuum, to perform the four-probe resistivity measurements, and topped by a $300 \: \mu m$-thick polyether-ether-ketone (PEEK) dome to perform XRD measurements in reflection geometry.

\subsection{XRD measurements}

The results on the lattice evolution under uniaxial tensile stress were obtained from XRD measurements performed with a $8 \: keV$ X-ray beam generated by a copper anode source (Rigaku RU-300B). The source is equipped with multilayer monochromator only allowing $K_{\alpha 1}$ and $K_{\alpha 2}$ emissions. The sample, in the biaxial tensile stress device, is positioned at the center of rotation of a Eulerian 4-circle diffractometer and probed in reflection geometry. Diffracted beams are detected with a 2D detector located $82 \: cm$ behind the sample. 3 non-collinear Bragg reflections (($0 \:\:\: 16 \:\:\: 0$), ($1 \:\:\: 15 \:\:\: 0$) and ($0 \:\:\: 16 \:\:\: 1$)) were recorded on the detector and projected along the $q_x$, $q_y$, $q_z$ directions of the reciprocal space and $2\theta$ to compute the lattice parameters $\alpha$, $\beta$, $\gamma$, $a$, $b$ and $c$ for each set of applied forces. The measurements were done on sample 1 under tensile stress along the c-axis at several temperatures ($250 \: K, 280 \: K, 310 \: K, 340 \: K, 370 \: K$) and extrapolated to tensile stress along a-axis regarding the linear dependance between the $a/c$-ratio and the tensile stresses in $TbTe_3$ \cite{gallofrantz_charge_2024}.

\subsection{Transport measurements}

To probe the electrical properties in the samples, four contacts were deposited in four-bar geometry along the c- and along the a-axis for sample 1 and sample 2, respectively. The sample 1 and the four-bar geometry are shown in Fig.\ref{fig:TbTe3_sample1_picture}. Keithley 2611 Sourcemeter and Keithley 2182a Nanovoltmeter were used to apply the current and to measure the voltage respectively. To measure the evolution of the samples electronic properties as a function of temperature and uniaxial tensile stresses, the forces $F_c$ and $F_a$ were changed at $350 \: K$, the resistances were measured during cooling down to $250 \: K$ at a fixed rate of $1 \: K/min$ and $V(I)$ curves are measured from $250 \: K$ at fixed temperature by heating in $5 \: K$ steps until $350 \: K$. We followed the same protocol for all measurements. It ensures to avoid sample aging and time effect on sliding properties \cite{frolov_features_2019, frolov_toward_2020, frolov_non-equilibrium_2021, frolov_logarithmic_2023}.

\section{\label{sec:Experimental_results} Experimental results}
\subsection{Correlation between the lattice and the CDWs properties under tensile stress}

\begin{figure}[h]
    \centering
    \includegraphics[width = 0.85\linewidth]{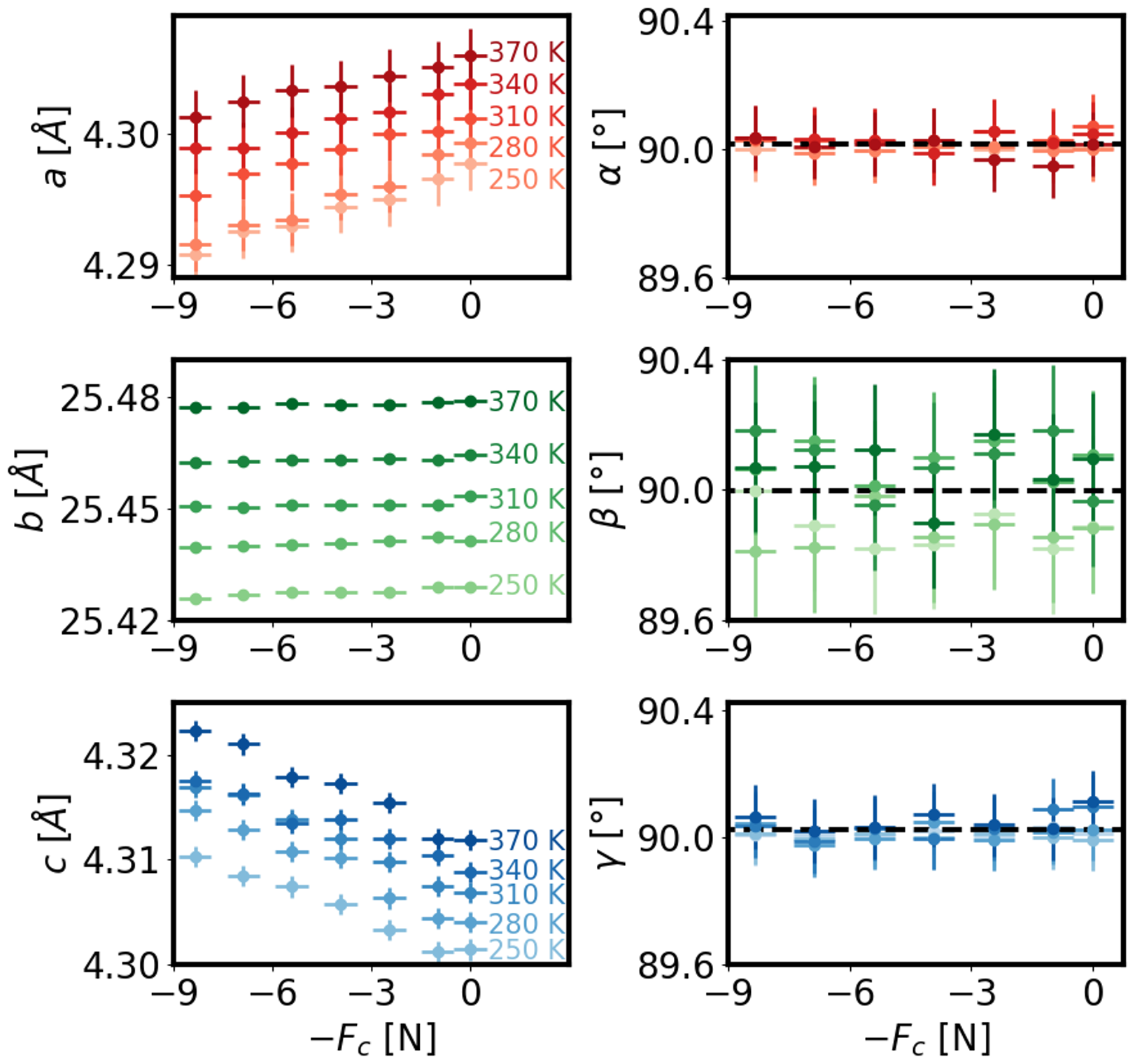}
    \caption{Evolution of lattice constants $TbTe_{3}$ (sample 1) obtained from XRD measurement at several temperatures.}
    \label{fig:lattice_evolution_under_stress}
\end{figure}

As detailed in the experimental methods (Sec.\ref{sec:Experimental_methods}.C), we measured the rocking curves around three non-collinear Bragg reflections (($0 \:\:\: 16 \:\:\: 0$), ($1 \:\:\: 15 \:\:\: 0$) and ($0 \:\:\: 16 \:\:\: 1$)) to follow the evolution of all lattice parameters under uniaxial tensile stress along c. The evolutions of all lattice constants $a, b, c, \alpha, \beta, \gamma$ are shown as function of $-F_c$, for visual continuity reasons, in Fig.\ref{fig:lattice_evolution_under_stress}. The lattice constants a and c display a linear dependance with applied force at all temperature with an in-plane Poisson ration $\nu_{ac} = -\varepsilon_{aa} / \varepsilon_{cc} \sim 1$, as already reported in \cite{gallofrantz_charge_2024}. When the tensile stress is applied along the c-axis, c increases, a decreases and b decreases too with an out-of-plane Poisson ration $\nu_{bc} = -\varepsilon_{bb} / \varepsilon_{cc} \sim 0.1$ for all temperatures, in agreement with the value reported in \cite{gallofrantz_charge_2024}.

\begin{figure}[h]
    \centering
    \includegraphics[width = 0.70\linewidth]{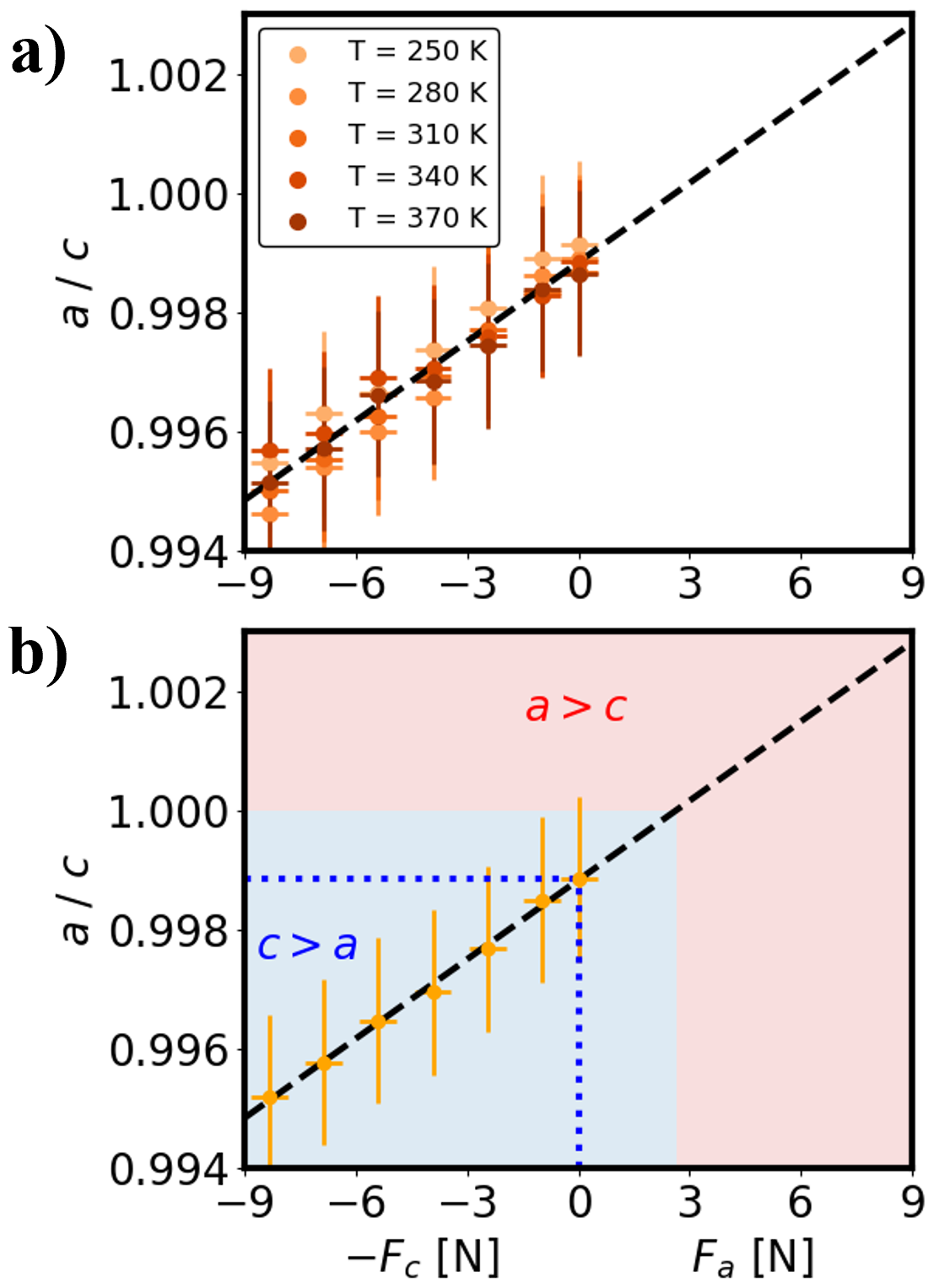}
    \caption{Evolution of the $a/c$-ratio as a function of tensile stress along the c-axis a) for several temperatures and b) averaged over the temperatures For both, the ratio has been extrapolated for tensile stress applied along a-axis ($F_a$).}
    \label{fig:a_over_c_under_stress}
\end{figure}

\noindent This order of magnitude difference between $\nu_{ac}$ and $\nu_{bc}$ can be attributed to the Van-der-Waals coupling along the b-axis, weaker than the in-plane coupling. Similarly to \cite{gallofrantz_charge_2024}, the weak variation of the angles $\alpha$, $\beta$ and $\gamma$ as a function of the applied force allows ($< 0.5\%$) to neglect the shear components of the strain tensor. For all temperatures, the a/c-ratio is computed and its variation as a function of the applied force along c-axis $F_c$ is illustrated in Fig.\ref{fig:a_over_c_under_stress}.\\

\newpage
\noindent As already reported in Ref.\cite{gallofrantz_charge_2024} and illustrated in Fig.\ref{fig:a_over_c_under_stress}a), the $a/c$-ratio displays a linear dependance with the respect to the applied force, regardless of the temperature. The linear dependance of the ratio, averaged over the temperature and shown in Fig.\ref{fig:a_over_c_under_stress}b), allows to extrapolate the ratio for tensile stress applied along $a$-axis ($F_a$), crossing the unity ($F_a \sim 2.6 \: N$) and reaching values greater than one ($a > c$). Previously, it has been shown that the inversion of this ratio is accompanied by an orientation transition from $c$-axis to $a$-axis for the CDW with a coexistence of region close $a/c \sim 1$ \cite{gallofrantz_charge_2024}. This transition was identified using transport measurements in both samples, and all relevant quantities will be hereafter expressed using the $a/c$-ratio.

\begin{figure}[h]
    \centering
    \includegraphics[width = 1.00\linewidth]{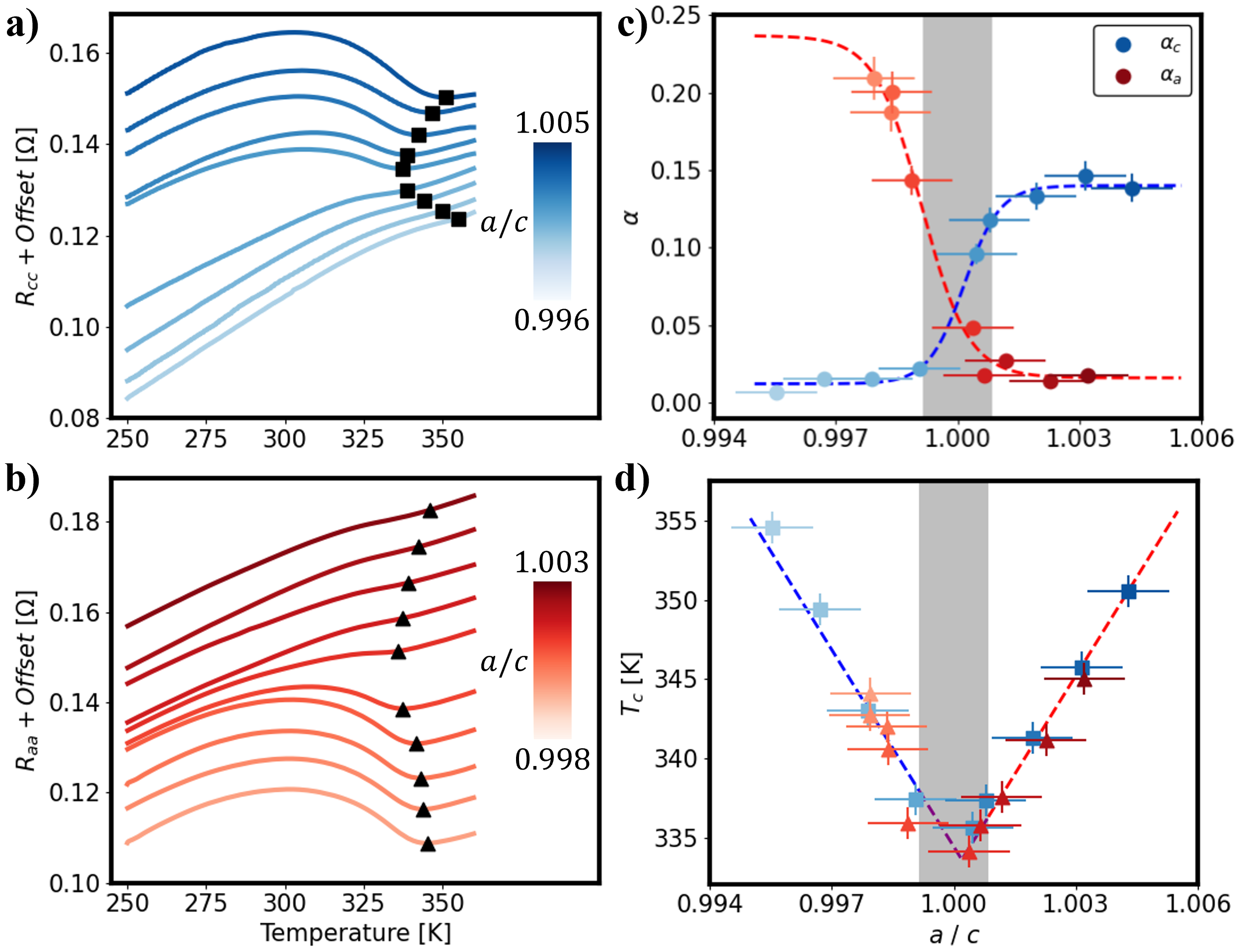}
    \caption{Resistances variation as a function temperature and $a/c$-ratio for a) sample 1 ($R_{cc}$, blue curves) and b) sample 2 ($R_{aa}$, red curves). The curves are shifted with an offset for clarity. The critical temperature is extracted from the maximum of second derivative and shown on each curve with black square in a) and black triangle in b). c) $\alpha_a$ and $\alpha_c$ and d) $T_c$ variations as a function the $a/c$-ratio. For both panels, $a/c\sim 1 \pm 0.001$ is highlighted by a dark rectangle (Blue and red dots from $R_{cc}$ and $R_{aa}$ respectively).}
    \label{fig:Rcc_Raa_properties}
\end{figure}

\noindent As detailed in the experimental methods (Sec.\ref{sec:Experimental_methods}.D), tensile stress and temperature dependance of the resistance has been measured in four bars geometry for both samples. Resistance along $c$-axis was probed in sample 1 (see Fig.\ref{fig:TbTe3_sample1_picture}) while resistance along $a$-axis was probed in sample 2. Both evolutions are illustrated as a function of $a/c$-ratio in Fig.\ref{fig:Rcc_Raa_properties}a) and b), respectively. The CDW critical temperature $T_c$ is extracted from the local maximum of the second derivative of the resistance curves.

\noindent The parameter $\alpha$ measures the size of the resistive increment. Using a simple Drude approach, this increment results from the competition between the diminution of the charge carriers density and the increase of the mean free time because of the gap opening. The parameter is usually defined as $\alpha_i = (R_i - R_{M, i})/R_{M, i}$ with $i = (a, c)$ and where $R_{i}$ is the resistance at the local maximum of $R(T)$ below the CDW transition. $R_{M, i}$ is the value of $R_M(T)$ at the same temperature as $R_i$, and $R_M(T)$ is the linear curve extrapolated in the full temperature
range from the metallic part of $R(T)$, above T$_c$ (see Supplementary information of Ref.\cite{gallofrantz_charge_2024} and \cite{gallofrantz_tuning_2026}). $\alpha_c$ measures increment of the resistance probed along $c$-axis in sample 1, while $\alpha_a$ measures the one probed along $a$-axis in sample 2. In these 2D systems, due to the distribution of Fermi velocities, an increase of $\alpha_a$ (respectively $\alpha_c$) is the result of gap opening due to a CDW formation in the perpendicular direction $CDW_c$ (respectively $CDW_a$) \cite{sinchenko_spontaneous_2014}. The variations of $\alpha_a$, $\alpha_c$ and $T_c$ as a function of $a/c$-ratio are displayed in Fig.\ref{fig:Rcc_Raa_properties}c) and d) respectively. Their behaviour is the same as the one probed in Montgomery geometry reported in ref.\cite{gallofrantz_charge_2024}.\\

\noindent In Fig.\ref{fig:Rcc_Raa_properties}c), both $\alpha_a$ and $\alpha_c$ variations are fitted with sigmoid curves. $\alpha_a$ (red dots and red dashed curve) saturates around $0.23$ for $a/c < 1$ showing the existence of the $CDW_c$ in the crystal. While the $a/c$-ratio increases up to $a/c > 1$, $\alpha_a$ decreases to almost zero indicating the disappearance of the $CDW_c$. Simultaneously, $\alpha_c$ (blue dots and blue dashed curve) increases from $0$ for $a/c < 1$ to $0.14$ and saturates for $a/c > 1$ revealing the appearance of the perpendicular charge order $CDW_a$. On the other hand, the critical temperature $T_c$ demonstrates a linear behaviour with $|a/c - 1|$ with a minimum for $a/c \sim 1$ and without saturation regimes for $|a/c - 1| > 0$. All these variations have been previously reported in Ref.\cite{gallofrantz_charge_2024} and show the sensitivity of both samples to tensile stress.

\newpage
\subsection{Temperature-stress evolution of CDWs sliding properties}

\begin{figure}[h]
    \centering
    \includegraphics[width = 1.00\linewidth]{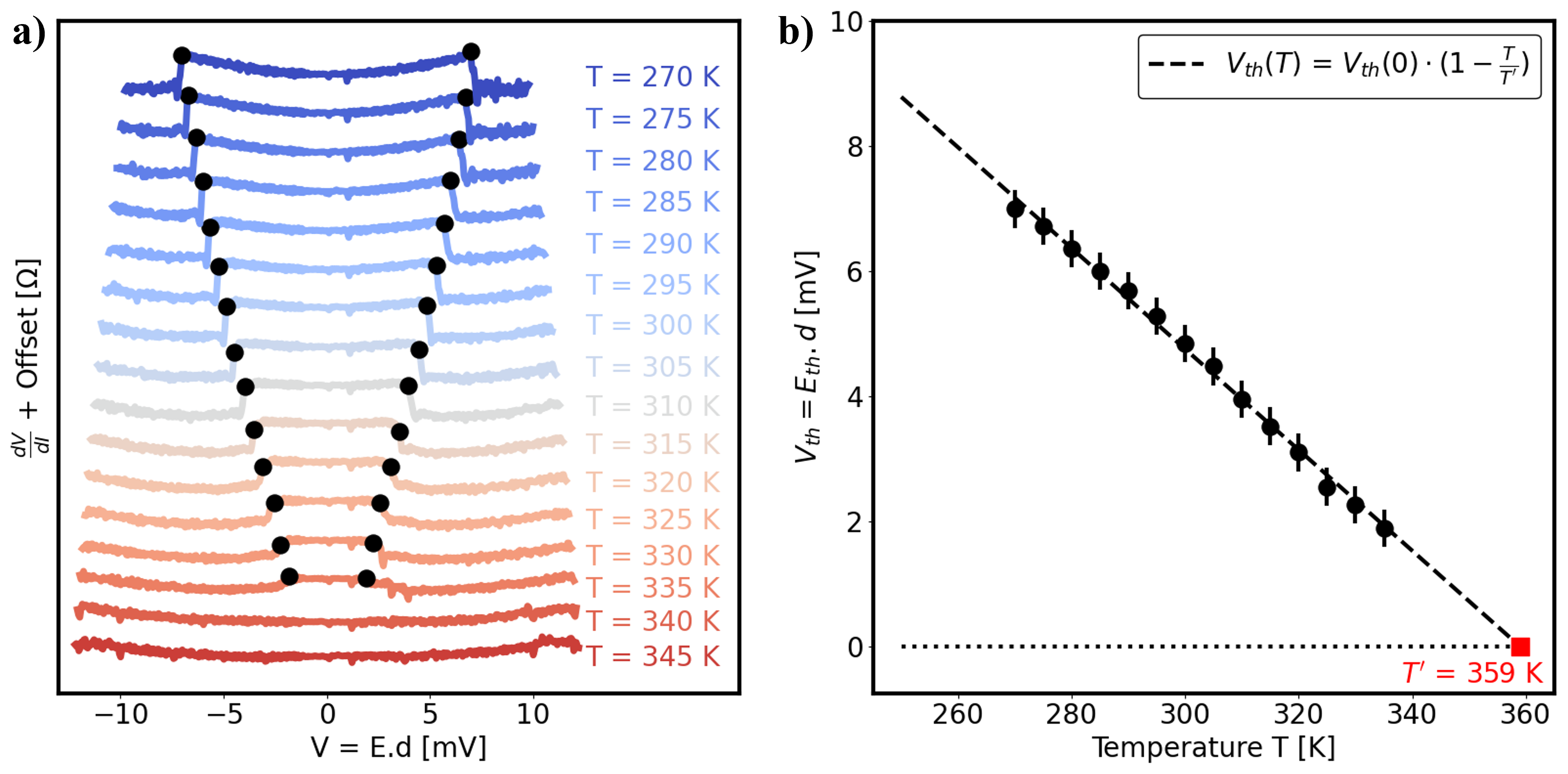}
    \caption{a) Differential resistances measured along $\vec{c}$ as a function of the internal voltage for several temperatures in the pristine $CDW_c$ phase of sample 1. The curves are shifted for clarity and the black spots highlight the threshold voltage $V_{th}$ for each temperature. b) Temperature evolution of the threshold voltage $V_{th}(T)$. Black dots are the experimental data extracted from a) while the dashed line corresponds to the fit following Eq.\ref{eq:temperature_dependence_threshold_voltage}.}
    \label{fig:temperature_dependance_sliding}
\end{figure}

\noindent As mentioned in the experimental methods (Sec.\ref{sec:Experimental_methods}.D), the non-linear transport properties were probed for each CDW phase as a function of temperature and tensile stress. $CDW_c$ and $CDW_a$ properties were measured in sample 1 and 2 respectively. For each value of the $a/c$ ratio, the V(I) curves were measured from $250 \: K$ to $350 \: K$. The temperature evolution measured in the pristine $CDW_{c}$ state between $270 \: K$ and $345 \: K$  is shown in Fig.\ref{fig:temperature_dependance_sliding} a) and b). The differential resistances are plotted as a function of internal voltage for several temperatures in Fig.\ref{fig:temperature_dependance_sliding}a). For each curve, the drop of resistance allows to identify the threshold voltage $V_{th}$ related to the threshold electric field by $V_{th} = E_{th} \times d$ where $d$ is the distance between the two electrodes $V^{+}$ and $V^{-}$ (see Fig.\ref{fig:TbTe3_sample1_picture}).\\

\noindent The threshold voltage is obtained for each temperature by averaging the positive and negative threshold voltages, highlighted with black dots on the $dV/dI$ curves shown in  Fig.\ref{fig:temperature_dependance_sliding}a). The temperature dependence of the threshold voltages obtained in the pristine $CDW_c$ state is shown in Fig.\ref{fig:temperature_dependance_sliding}b). The threshold voltage decreases linearly as temperature increases in the probed temperature range. $V_{th}$ can thus be described by the following formula \cite{sinchenko_unidirectional_2014} :

\begin{equation}
    V_{th}(T) = V_{th}(0). \left( 1 - \frac{T}{T'} \right)
    \tag{III.B.1}
    \label{eq:temperature_dependence_threshold_voltage}
\end{equation}\\

\noindent where $V_{th}(0) = 24.6 \pm 0.4 \: mV$ and $T' = 359 \pm 1 \: K \approx (1.07 \pm 0.01) \times T_c$ for the pristine state, and $V_{th}(T_c) \neq 0$, in agreement with the values reported in the literature \cite{sinchenko_unidirectional_2014}. Interestingly, the extrapolated temperature $T' = 359 \pm 1 \:K$ is close to the temperature at which the CDW signal fully disappears in X-ray diffraction experiments in TbTe$_3$ (363 K reported in \cite{ru_effect_2008}). Between $T_c$ and this temperature, a broad and weak scattering is observed, accounting for a CDW with short-range correlations lengths. 

\begin{figure}[h]
    \centering
    \includegraphics[width = 0.95\linewidth]{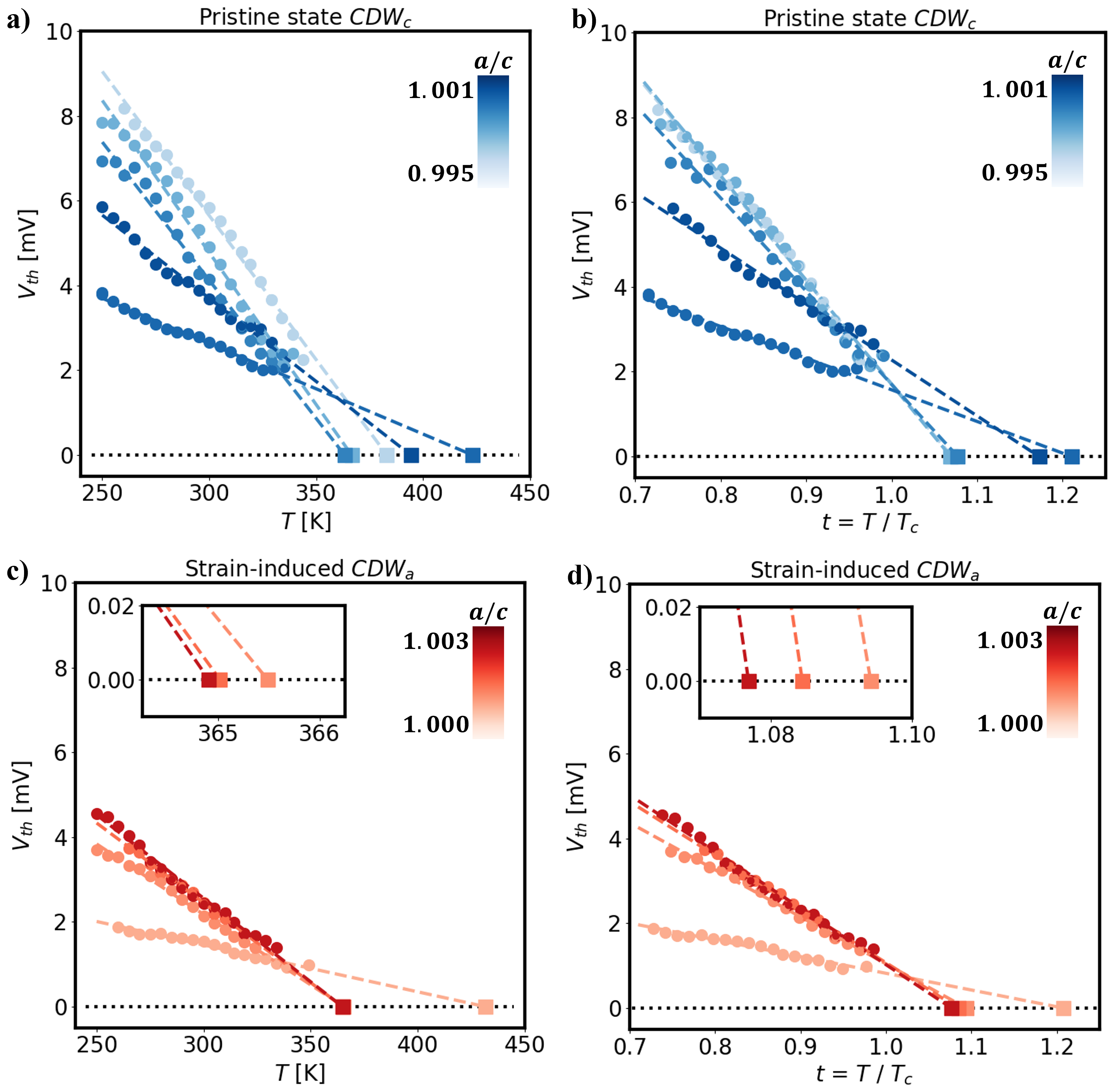}
    \caption{Temperature and reduced temperature variations of the threshold voltages of the pristine state $CDW_c$ (a) and b)) and the strain-induced $CDW_a$ (c) and d)) for different $a/c$-ratio. Fits with Eq.\ref{eq:temperature_dependence_threshold_voltage} are shown in dashed lines with corresponding colours in both panels.}
    \label{fig:temperature_stress_dependance_sliding}
\end{figure}

\noindent We then performed the same series of measurements, at different temperatures, for different strain states, and thus different $a/c$ ratios, to explore the evolution of the sliding properties of $CDW_c$ (resp. $CDW_a$), by measuring $V(I)$ curves and get $dV/dI$ along $\vec{c}$ (resp. $\vec{a}$) in sample 1 (resp. sample 2). Note that as $T_c$ significantly changes with $a/c$, it is necessary to use the normalized temperatures $t = T/T_c$ as meaningful parameter when comparing data obtained in different strain states. We then use the same analysis procedure as in the pristine state to get the $a/c$ dependence of $V_{th}(T)$ and $V_{th}(t = T/T_c)$, shown in Fig.\ref{fig:temperature_stress_dependance_sliding}a) and b) for $CDW_c$, and in Fig.\ref{fig:temperature_stress_dependance_sliding}c) and d) for $CDW_a$. \\

\noindent For all values of $a/c$ and for both $CDW$ orders, $V_{th}$ decreases linearly while the temperature increases. Moreover, for both orders, far from the coexistence phase ($a/c \ll 1$ for $CDW_c$ and $a/c \gg 1$ for $CDW_a$) the slopes of $V_{th}(T)$ are similar. However, when the $a/c$-ratio is close to 1, i.e. when the crystal exhibits a coexistence of $CDW_c$ and $CDW_a$, the slopes of $V_{th}(T)$ are lower for $CDW_c$ and $CDW_a$, as illustrated in Fig.\ref{fig:temperature_stress_dependance_sliding}a) and c), respectively. Remarkably, the finite value $V_{th}(T_c)$ is independent on $a/c$: all data points fall at the same threshold voltage when $t=1$ in Fig.\ref{fig:temperature_stress_dependance_sliding}b) and d). \\

\begin{figure}[h]
    \centering
    \includegraphics[width = 1.00\linewidth]{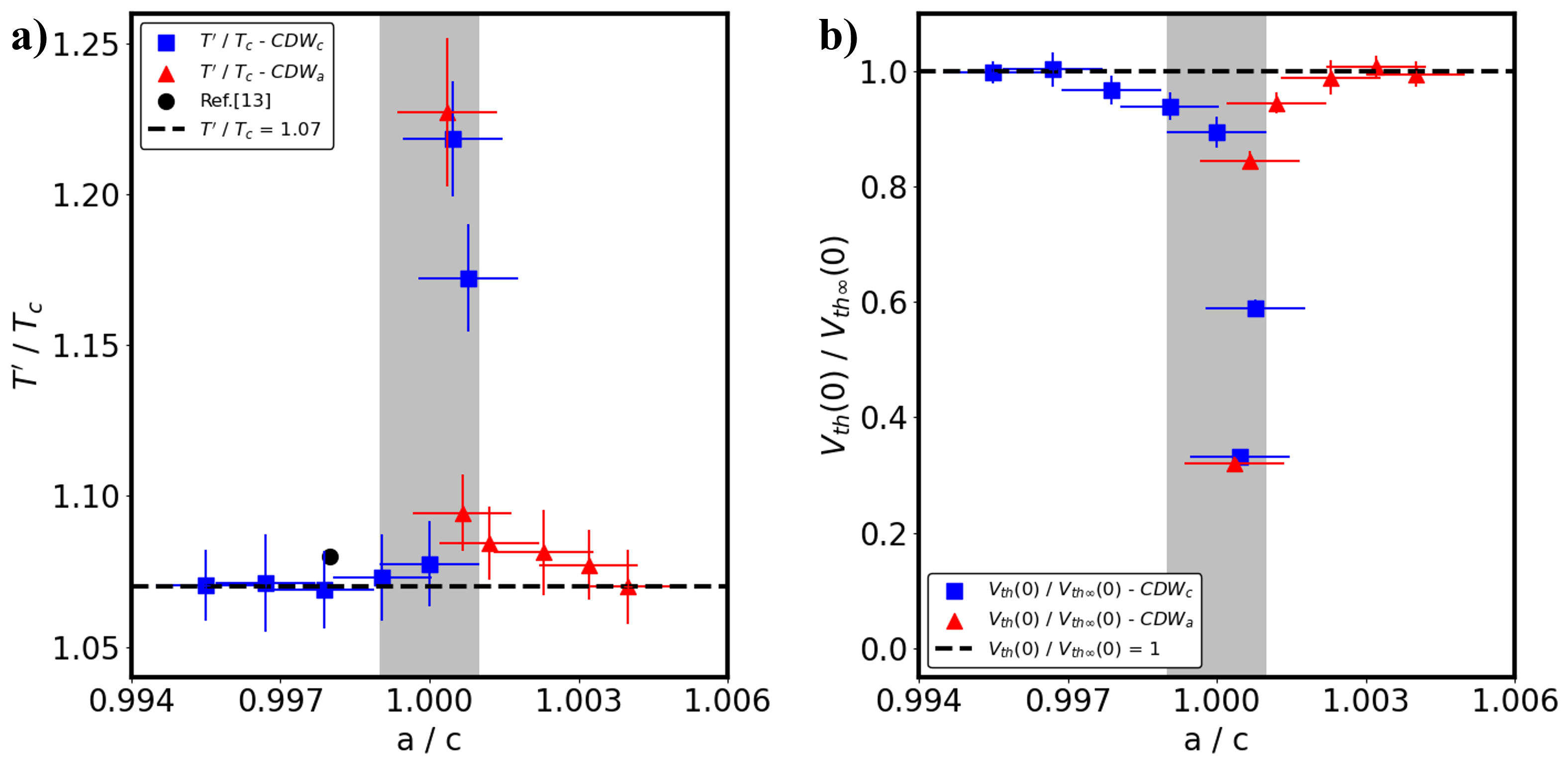}
    \caption{a) $T'/T_c$ and b) $V_{th} / V_{th \infty}$ as a function of $a/c$ for $CDW_c$ (sample 1, blue square) and $CDW_a$ (sample 2, red triangles). The $T'/T_c$ value obtained in the pristine state in Ref.[13] is illustrated with a black dot. The dashed lines highlight the limit values $T'/T_c = 1.07$ in a) and $V_{th} / V_{th \infty} = 1$ in b). For a) and b), the coexistence region is highlighted by a grey region centered in $a/c = 1$ with a width of $\pm 0.001$.}
    \label{fig:fit_parameters_variations_aoverc}
\end{figure}

\newpage
\noindent For both $CDWs$, the data are fitted with Eq.\ref{eq:temperature_dependence_threshold_voltage} and the variations of the meaningful parameters $T' / T_c$ and $V_{th}(0) / V_{th \infty}(0)$ are shown in Fig.\ref{fig:fit_parameters_variations_aoverc} as a function of $a/c$, where $V_{th \infty}(0)$ is the threshold voltage taken far from the coexistence region ($V_{th \infty}(0) = V_{th}(T = 0, \: a/c \ll 1)$ for $CDW_c$ and $V_{th \infty}(0) = V_{th}(T = 0, \: a/c  \gg 1)$ for $CDW_a$). The $a/c$ dependence of $T'/T_c$ is shown in Fig.\ref{fig:fit_parameters_variations_aoverc}a). Far from the coexistence region, it displays a common limit value for both $CDW$ orders : $T'/T_c \approx 1.07 \pm 0.01$, but it displays a sharp divergence when $a/c \sim 1$. A similar behaviour is observed for $V_{th}(0) / V_{th \infty}(0)$ in Fig.\ref{fig:fit_parameters_variations_aoverc}b) : far from the coexistence $V_{th}(0)$ saturates while it dramatically drops when $a/c \sim 1$. 

\section{Discussion}

In $TbTe_3$ under tensile stress applied along $\vec{a}$, the pristine $CDW_c$ state disappears and a new $CDW_a$ order appears along the perpendicular $a$-axis  \cite{gallofrantz_charge_2024}. Our measurements of the non-linear transport properties demonstrate that the strain-induced $CDW_a$ can slide like the pristine $CDW_c$, highlighting its incommensurability with the host lattice. Moreover, for both $CDW$ orders, the threshold field continues to show a linear temperature dependence with similar valuers and slopes, suggesting that they may correspond to symmetry-related manifestations of a same underlying order. This common behaviour can be understood in terms of a restored symmetry connecting the two $CDW$ states \cite{kim_emergent_2024}. Nevertheless, the slope of $V_{th}(T)$ is dramatically reduced in the coexistence phase ($a/c \sim 1$), suggesting that the sliding is enhanced when both CDW orders coexist in the crystal.\\

\noindent The extrapolated temperature $T'$ at which $V_{th}(T)$ indeed reaches zero closely coincides with the upper temperature limit of the short-range CDW order observed by XRD \cite{ru_effect_2008}. Above $T'$, no distinct CDW peaks could be resolved experimentally, suggesting that CDW correlations become too small to be detected. Upon cooling below $T'$ and above T$_c$, short-range CDW correlations progressively develop in the crystal without establishing a long-range order. In this intermediate regime, ($T_c < T < T'$), the absence of a measurable threshold voltage may indicate that these correlations are not sufficiently spatially coherent for collective CDW sliding. Then, below T$_c$, the development of a long-range CDW order provides the spatial coherence required for the appearance of sliding above a finite threshold. Moreover, close to T$_c$, the CDW condensate starts to develop, and the finite value of $E_{th}(T_c)$ may reflect pinning of the emerging CDW by intrinsic defects. Below T$_c$, the amplitude of the CDW condensate increases and modifies its interaction with the pinning, leading to an increase of the threshold field. Then, the close correspondence between $T'$ and the upper temperature limit of the short-range CDW order observed by XRD suggests that the temperature dependence of the threshold voltage may provide an indirect estimate of the onset of loss of long-range CDW order. Hence the dimensionless quantity $T'/T_c - 1$ provides a measurement of the relative range of the short-range CDW order above $T_c$ (see Fig.\ref{fig:fit_parameters_variations_aoverc}a)).

\noindent Notably, this quantity increases from approximately $0.07$ in the single-CDW state (pure $CDW_c$ or pure $CDW_a$) to $0.22$ in the coexistence phase. In the latter, the lattice shows a square geometry, making the two perpendicular crystallographic directions $\textbf{a}$ and $\textbf{c}$ nearly equivalent. This isotropy makes the two CDW orientations equivalent, which may allow fluctuations between them to persist over a wider temperature range. Moreover, $T_c$ reaches a minimum in this regime (see Fig.\ref{fig:Rcc_Raa_properties} c)), suggesting that the higher symmetry between the two perpendicular directions reduces the tendency to form a long-range $CDW$ state. \\

\noindent Within this picture, the linear temperature dependence of the threshold electric field observed here is rather unusual in the context of defect or surface pinnings. However, similar behaviours have been reported in other compounds of higher dimensionality, in particular NbS$_3$ \cite{semakin_features_2024} and $1T$-TiSe$_2$ \cite{wei_depinning_2024}. In contrast, in quasi-1D CDW systems such as NbSe$_3$, the threshold field exhibits a divergence at low temperature and close to $T_c$ \cite{fleming_electric-field_1980}. This behaviour was experimentally observed and described in terms of thermal fluctuations of the CDW phase and the dimensionality of the CDW \cite{maki_thermal_1986, maki_impurity_1989, maki_phase_1990}.\\

\noindent However, in quasi-1D systems, a linear temperature dependence of the threshold field can be recovered by considering the effect of a commensurability potential. Following the theoretical framework of Ref.\cite{maki_phase_1990}, a fourth-order commensurability potential can be introduced and written as :
\begin{equation}
    V_{c}(T) \propto \Delta(T)^4.cos(4\phi(x))
    \tag{IV.1}
    \label{eq:commensurability_potential}
\end{equation}

\newpage
\noindent where $\Delta(T)$ is the temperature-dependent order parameter and $\phi(x)$ is the spatially varying CDW phase. The corresponding threshold electric field is given by :
\begin{equation}
    \frac{E_{th}^{c}(T)}{E_{th}^{c}(0)} = e^{-16T/T_0} .\left( \frac{\Delta(T)}{\Delta_0} \right)^{4}.\frac{n_s(0)}{n_s(T)}
    \tag{IV.2}
    \label{eq:commensurate_threshold_field}
\end{equation}\\

\noindent where $E_{th}^{c}(T)$ is the contribution of the commensurability potential to the threshold field, $T_0$ is a characteristic temperature associated with the thermal phase fluctuations, and $n_s$ denotes the condensate density. Regarding Ref.\cite{maki_phase_1990}, the fourth-order commensurability potential leads to a linear decrease of the threshold field as the temperature increases. For a quarter-filled system ($2k_{F} \sim 1/4$), such that the fourth-order term satisfies the commensurability condition $4\times 2k_{F} \sim 1$, $E_{th}^{c}(T) \propto T_c - T$ with $E_{th}^{c}(T_c) = 0$ in the vicinity of T$_c$. Interestingly, the fourth-order commensurability potential is the only order for which the threshold field displays this linear behaviour.\\

\noindent In TbTe$_3$, $2k_F \sim 5/7 c^*$, while the complementary wave vector can be defined as $q^{*} = 1 - 2k_F \sim 2/7 c^*$. Therefore, the strict quarter-filling commensurability condition is not satisfied. Nevertheless, the linear decrease of the threshold field with increasing temperature is consistent with the behaviour predicted for the fourth-order commensurability potential. However, for all values of $a/c$-ratio, the threshold fields do not extrapolate to zero at T$_c$ (see Fig.\ref{fig:temperature_stress_dependance_sliding}). This residual threshold is attributed to pinning by crystal defects $E_{th}^{pin}$, which is temperature and tensile stress independent. The total threshold field can therefore be expressed phenomenologically as :

\begin{equation}
    \begin{split}
        E_{th}(T) &= E_{th}^{c}(T) + E_{th}^{pin} \\\\
        &= E_{th}^{c}(0).e^{-16T/T_0} .\left( \frac{\Delta(T)}{\Delta_0} \right)^{4}.\frac{n_s(0)}{n_s(T)} + E_{th}^{pin}
    \end{split}
    \tag{IV.3}
    \label{eq:total_threshold_field}
\end{equation}\\

\noindent With this description, all curves in Fig.\ref{fig:temperature_stress_dependance_sliding}b) and d) are expected to intersect at $t = T/T_c = 1$. Moreover, the different pinning contributions and the CDW fluctuations temperature range $T' / T_c - 1$ are linked to the thermal fluctuations of the phase through the following relation (see App.~\ref{app:expansion_electrical_threshold_field}).

\begin{equation}
    \frac{E_{th}^{pin}}{E_{th}^{c}(0)} \approx 4.692 \: \left( \frac{T'}{T_c} - 1 \right) .e^{-16T_c/T_0}
    \tag{IV.4}
    \label{eq:link_pinning_fluctuations}
\end{equation}\\

\begin{figure}[h]
    \centering
    \includegraphics[width = 1.00\linewidth]{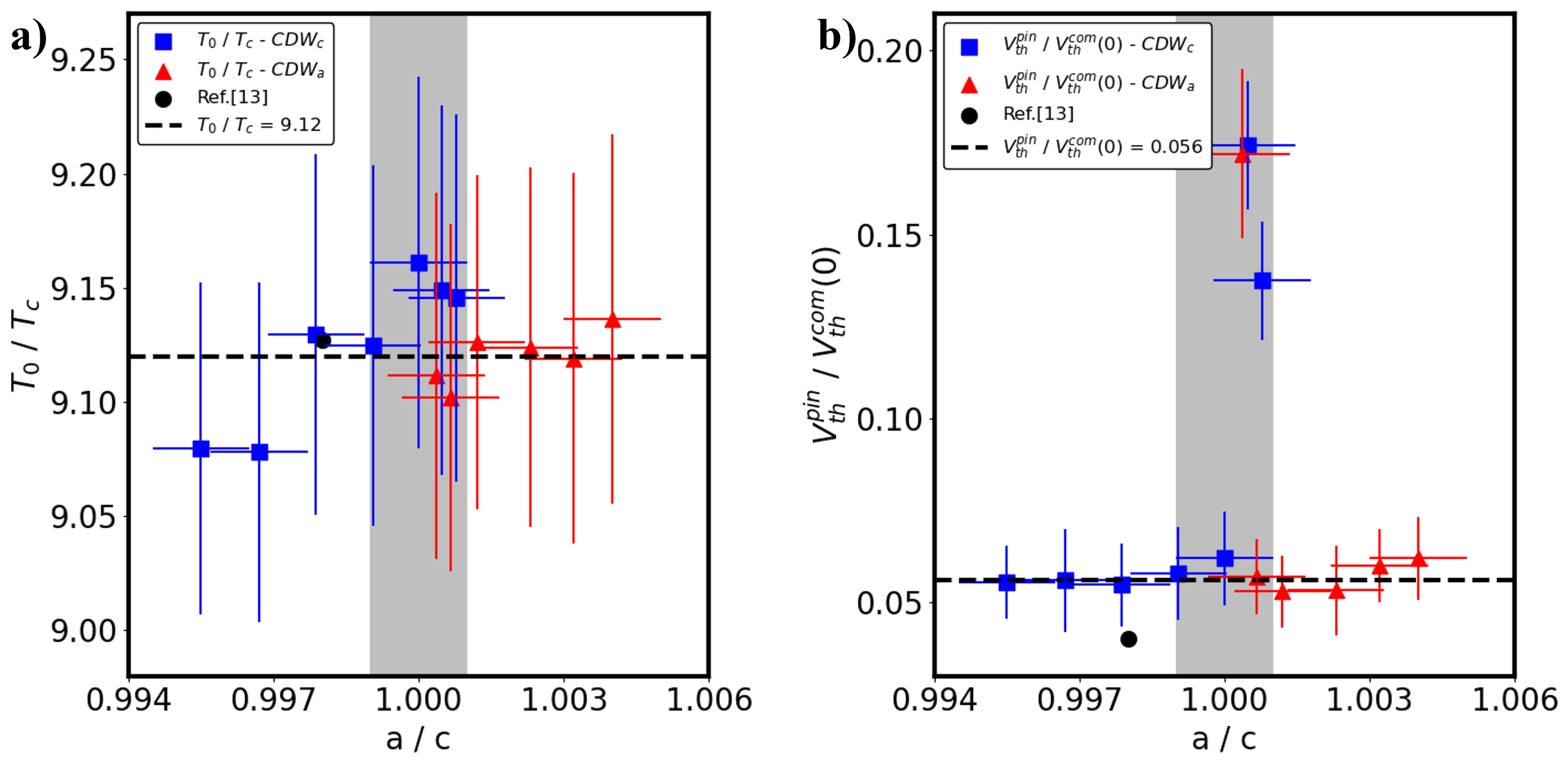}
    \caption{a) $T_0/T_c$ and b) $E_{th}^{pin} / E_{th}^{c}(0)$ as a function of $a/c$-ratio for $CDW_c$ (blue square) and $CDW_a$ (red triangles). Pristine state taken from Ref.\cite{sinchenko_unidirectional_2014} is illustrated with a black dot for both. Dashed lines highlight $T_0 /T_c = 9.12$ in a) and $E_{th}^{pin} / E_{th}^{c}(0) = 0.056$ in b). For a) and b), the coexistence region is highlight by a dark rectangle centered in $a/c = 1$ with a width of $\pm 0.001$.}
    \label{fig:fit_com_aoverc}
\end{figure}

\noindent For each curve shown in Fig.\ref{fig:temperature_stress_dependance_sliding}, we extrapolate $E_{th}^{pin}$ from the linear temperature dependence of $E_{th}(T)$ to $T = T_c$. The temperature dependences of the order parameter and condensate density do only depend on $T_c$ and are taken into account to extract the thermal fluctuations contribution $T_0$ from Eq.\ref{eq:total_threshold_field}. The variations of $T_0 / T_c$ and $E_{th}^{pin}/E_{th}^{c}(0) = V_{th}^{pin}/V_{th}^{c}(0)$ are displayed in Fig.\ref{fig:fit_com_aoverc}a) and b) respectively as a function of $a/c$.

\noindent As shown in Fig.\ref{fig:fit_com_aoverc}, $T_0 / T_c$ does not exhibit clear variations as a function of $a/c$ and remains close to a mean value $T_0 / T_c \approx 9.12$. This indicates that the thermal phase-fluctuation contribution is essentially independent on lattice anisotropy. On the other hand, the ratio $E_{th}^{pin} / E_{th}^{c}(0)$ is constant out of the CDW coexistence phase while it diverges for both orders at $a/c \sim 1$. Since $T_0 / T_c$ remains approximately constant, the increase of $T'/T_c$ (and so of the temperature range of the short-range CDW order) leads to an increase of $E_{th}^{pin} / E_{th}^{c}(0)$. As $E_{th}^{pin}$ is independent of both temperature and tensile stress, this increase corresponds to a lower commensurability contribution $E_{th}^{c}(0)$ in the coexistence region.\\ 

\noindent The reduction of $E_{th}(0)$ indicates an enhanced sliding response of both CDW orders in the coexistence regime. This behaviour is particularly relevant because, as the lattice approaches the nearly isotropic limit $a/c \sim 1$, the two perpendicular crystallographic directions become nearly equivalent by symmetry. The crystal therefore allows both CDW orders to develop and the short-range CDW order to persist over a larger temperature range. This behaviour contrasts with the expectation that the coexistence of two perpendicular CDW orders might enhance their mutual locking or pinning. Instead, the appearance of in-plane symmetry is accompanied by a reduction of the commensurability contribution and an easier sliding of both CDW orders.\\

\noindent The fourth-order commensurability potential considered in Ref.\cite{maki_phase_1990} is derived for 1D CDW system satisfying the commensurability condition $(4 \times k_F \sim 1)$ although this condition is not strictly fulfilled in the present compound. Moreover, the physical meaning of the characteristic temperature $T_0$ is not established in the present case. In quasi-1D compounds, $T_0$ is related to the electronic hopping parameters normalized by the CDW gap, whereas its microscopic origin and relevance to CDW sliding in a two-dimensional system remains unclear. 

However, our measurements highlight the incommensurability of the strain-induced $CDW_a$ with the host lattice. In addition, they demonstrate as a clear link between the sliding properties of the $CDWs$ stabilized in TbTe$_3$ and the in-plane strain state of this quasi-2D crystal, experimentally established by a combination of structural and non-linear transport measurements. We show in particular that the sliding of both in-plane $CDWs$ is enhanced in the strain-induced coexistence state of these orthogonal $CDWs$. This work raises questions about how the pristine $CDW_c$ and the strain-induced $CDW_a$ coexist and interact within the crystal at microstructural level, which is necessary to build a microscopic description of their coupling in such a two-dimensional system.


\newpage
\appendix
\begin{widetext}
\section{\label{app:expansion_electrical_threshold_field} Expansion of the Electrical threshold field in the vicinity of T$_c$}

We recall here the temperature evolution of the electrical threshold field defined in Eq.\ref{eq:total_threshold_field} :

\begin{equation}
    \begin{split}
        E_{th}(T) &= E_{th}^{c}(T) + E_{th}^{pin} \\
        &= E_{th}^{c}(0).e^{-16T/T_0} .\left( \frac{\Delta(T)}{\Delta_0} \right)^{4}.\frac{n_s(0)}{n_s(T)} + E_{th}^{pin}
    \end{split}
    \tag{A.1}
    \label{eq:app_total_threshold_field}
\end{equation}\\

\noindent In the vicinity of the critical temperature (T$_c$), the electrical threshold field can be written as follows :
\begin{equation}
    E_{th}(T \approx T_c) =  \left. \frac{dE_{th}(T)}{dT} \right|_{T_c} .(T - T_c)+ E_{th}(T_c)
    \tag{A.2}
    \label{eq:app_expansion_total_threshold_field_1}
\end{equation}\\

\noindent By definition :
\begin{equation}
    \begin{split}
        E_{th}(T = T_c) &= E_{th}^{c}(T_c) + E_{th}^{pin} \\
        &= 0 +  E_{th}^{pin}\\
        &= E_{th}^{pin}
    \end{split}
    \tag{A.3}
    \label{eq:app_expansion_total_threshold_field_2}
\end{equation}\\

\noindent One can show from Eq.\ref{eq:app_total_threshold_field} that :
\begin{equation}
    \begin{split}
        \frac{dE_{th}(T)}{dT} &= \frac{dE^{c}_{th}(T)}{dT}\\\\
        &= E^{c}_{th}(T) .\left[ -\frac{16}{T_0} + 4.\frac{\Delta'(T)}{\Delta(T)} - \frac{n_s'(T)}{n_s(T)} \right]
    \end{split}
    \tag{A.4}
    \label{eq:app_expansion_total_threshold_field_3}
\end{equation}\\

\noindent with $\Delta'(T) = d\Delta(T)/dT$, $n_s'(T) = dn_s(T)/dT$ and where $\Delta(T)$ and $n_s(T)$ are the temperature dependences of the order parameter and of the condensate density respectively. In the vicinity of T$_c$, the order parameter $\Delta(T)$ becomes :
\begin{equation}
    \begin{split}
        \Delta(T) &= \Delta_0 .tanh \left( 1.764 .\sqrt{\frac{T_c}{T} - 1}\right) \\\\
        &\sim  1.764\Delta_0 .\left(1 - \frac{T}{T_c}\right)^{1/2}\\\\
        \left( \frac{\Delta(T)}{\Delta_0} \right)^4 &= (1.764)^4 .\left(1 - \frac{T}{T_c}\right)^{2}
    \end{split}
    \tag{A.5}
    \label{eq:app_expansion_of_order_param}
\end{equation}\\

\newpage
\noindent And the superfluid density $n_s(T)$ can be written :
\begin{equation}
    \begin{split}
        \frac{n_s(T)}{n_s(0)} &= \frac{7.\zeta(3)}{4.\pi^2} .\left( \frac{\Delta(T)}{k_B .T_c} \right)^2 \\\\
        &= \frac{7.\zeta(3)}{4.\pi^2} .1.764^4 .\left(1 - \frac{T}{T_c}\right)
    \end{split}
    \tag{A.6}
    \label{eq:app_expansion_of_sf_density}
\end{equation}\\

\noindent Then, close to $T_c$, the electrical threshold component due the commensurability potential can written as follows :
\begin{equation}
    \begin{split}
        E^{c}_{th}(T) &= E_{th}^{c}(0) .e^{-16T/T_0} .\left( \frac{\Delta(T)}{\Delta_0} \right)^{4}.\frac{n_s(0)}{n_s(T)}\\\\
        &= E_{th}^{c}(0) .e^{-16T_c/T_0} . \frac{4 \pi^2}{7 \zeta(3)} .\left(1 - \frac{T}{T_c}\right)\\\\
        &= \frac{4 \pi^2}{7 \zeta(3)} .\frac{E_{th}^{c}(0)}{T_c} .e^{-16T_c/T_0} .(T_c - T)
    \end{split}
    \tag{A.7}
    \label{eq:app_expansion_of_electrical_threshold_fiels_part_comm}
\end{equation}\\

\noindent The derivatives $\Delta'(T)$ and $n_s'(T)$ can be expressed :
\begin{equation}
    \begin{split}
        \Delta'(T) &\sim -\frac{1.764\Delta_0}{2T_c}.(1 - \frac{T}{T_c})^{-1/2} \\\\
        \frac{\Delta'(T)}{\Delta(T)} &= \frac{-1/2}{T_c.(1 - T/T_c)} = \frac{-1/2}{T_c - T}\\\\
        n_s'(T) &\sim -\frac{n_s(0)}{T_c} .\frac{7.\zeta(3)}{4.\pi^2} .1.764^4 \\\\
        \frac{n_s'(T)}{n_s(T)} & = \frac{-1}{T_c .(1 - T/T_c)} = \frac{-1}{T_c - T} \\\\
    \end{split}
    \tag{A.8}
    \label{eq:app_expansion_of_derivatives}
\end{equation}

\noindent And then :
\begin{equation}
    4.\frac{\Delta'(T)}{\Delta(T)} - \frac{n_s'(T)}{n_s(T)} = \frac{-1}{T_c - T}
    \tag{A.9}
    \label{eq:app_expansion_of_order_param2}
\end{equation}\\

\noindent Combining Eq.\ref{eq:app_expansion_of_electrical_threshold_fiels_part_comm} and Eq.\ref{eq:app_expansion_of_order_param2}, Eq.\ref{eq:app_expansion_total_threshold_field_3} becomes :
\begin{equation}
    \begin{split}
        \frac{dE_{th}(T)}{dT} = \frac{4 \pi^2}{7 \zeta(3)} .\frac{E_{th}^{c}(0)}{T_c} .e^{-16T_c/T_0} .\left[-\frac{16}{T_0} .(T_c - T) - 1\right]
    \end{split}
    \tag{A.10}
    \label{eq:app_expansion_total_threshold_field_4}
\end{equation}

\noindent Finally Eq.\ref{eq:app_expansion_total_threshold_field_4} evaluated in $T_c$ becomes:
\begin{equation}
    \left. \frac{dE_{th}(T)}{dT} \right|_{T_c} = -\frac{4 \pi^2}{7 \zeta(3)} .\frac{E_{th}^{c}(0)}{T_c} .e^{-16 T_c/T_0}
    \tag{A.11}
    \label{eq:app_expansion_total_threshold_field_5}
\end{equation}\\

\noindent So Eq.\ref{eq:app_expansion_total_threshold_field_1} becomes :
\begin{equation}
    E_{th}(T \approx  T_c) = -\frac{4 \pi^2}{7 \zeta(3)} .\frac{E_{th}^{c}(0)}{T_c} .e^{-16 T_c/T_0} .(T - T_c) + E_{th}^{pin}
    \tag{A.12}
    \label{eq:app_expansion_total_threshold_field_6}
\end{equation}\\

\noindent In particular, at $T = T'$, we have :
\begin{equation}
    \begin{split}
        E_{th}(T = T') &= -\frac{4 \pi^2}{7 \zeta(3)} .E_{th}^{c}(0).e^{-16 T_c/T_0} .\frac{T' - T_c}{T_c}+ E_{th}^{pin}\\\\
        &= -\frac{4 \pi^2}{7 \zeta(3)} .E_{th}^{c}(0).e^{-16 T_c/T_0} .\left( \frac{T'}{T_c} - 1 \right) + E_{th}^{pin}\\\\
        &= 0
    \end{split}
    \tag{A.13}
    \label{eq:app_expansion_total_threshold_field_7}
\end{equation}\\

\noindent Resulting in :
\begin{equation}
    \boxed{
    \begin{split}
        \frac{T'}{T_c} - 1 &= \frac{7 \zeta(3)}{4 \pi^2}.\frac{E_{th}^{pin}}{E_{th}^{c}(0)}.e^{16T_c/T_0} \\\\
        &\approx  0.213 \: \frac{E_{th}^{pin}}{E_{th}^{c}(0)}.e^{16T_c/T_0} \\\\
        \frac{E_{th}^{pin}}{E_{th}^{c}(0)} &= \frac{4 \pi^2}{7 \zeta(3)} .\left( \frac{T'}{T_c} - 1 \right) .e^{-16T_c/T_0}\\\\
        &\approx 4.692 \: \left( \frac{T'}{T_c} - 1 \right) .e^{-16T_c/T_0}
    \end{split}
    }
    \tag{A.14}
    \label{eq:T_prime_as_a_function_of_other_parameters}
\end{equation}
\end{widetext}
\newpage
\nocite{*}
\newpage
\bibliography{TbTe3_tensile_stress_sliding.bib}
\end{document}